\documentclass[aoas,nameyear,dvips]{arximspdf}
\usepackage{graphics}
\doi{10.1214/09-AOAS302}
\volume{4}
\issue{2}
\pubyear{2010}
\firstpage{1056}
\lastpage{1080}

\makeatletter
\newcommand{\R}{\mathbb{R}}
\newcommand{\var}{\operatorname{var}}
\newcommand{\diag}{\operatorname{diag}}
\newcommand{\E}{\mathrm{E}}
\newcommand{\argmin}{\operatorname{argmin}}
\newcommand{\p}{\mathrm{P}}
\newcommand{\wt}[1]{\widetilde{#1}}
\newcommand{\wh}[1]{\widehat{#1}}
\newcommand{\eqref}[1]{(\ref{#1})}

\newcommand{\mc}[1]{\mathcal{#1}}
\newcommand{\T}{\top}
\newcommand{\sign}{\operatorname{sign}}
\newcommand{\indep}{\perp\!\!\!\perp}

\let\epsilon\varepsilon

\newcommand{\blanco}[1]{ }

\def\su{\sum_{i=1}^n}
\def\ninf{n \rightarrow\infty}

\newtheorem{theo}{Theorem}
\newtheorem{propo}{Theorem}

\newtheorem{definitioA}{Theorem}
\newtheorem{theoA}{Theorem}

\newtheorem{theorem}[theo]{Theorem}

\newtheorem{defnA}[definitioA]{Definition A.}
\newtheorem{theoremA}[theoA]{Theorem A.}
\newtheorem{prop}[propo]{Proposition}

\newcommand{\conv}[1]{\stackrel{\mathrm{#1}}{\rightarrow}}

\newcommand{\ik}{^{(k)}}

\makeatother

\begin{document}
\begin{frontmatter}

\title{Feature selection guided by structural information}
\runtitle{Feature selection guided by structural information}

\begin{aug}
\author[A]{\fnms{Martin} \snm{Slawski}\ead[label=e1]{ms@cs.uni-sb.de}\thanksref{t1,t2}\corref{}},
\author[B]{\fnms{Wolfgang} \snm{zu Castell}\ead[label=e2]{castell@helmholtz-muenchen.de}}
and
\author[C]{\fnms{Gerhard} \snm{Tutz}\ead[label=e3]{tutz@stat.uni-muenchen.de}}

\runauthor{M. Slawski, W. {\normalfont \textsc{zu}} Castell and G. Tutz}
\thankstext{t1}{A large fraction of this work was done while the
author was at
the Department of Statistics, University of Munich and the Sylvia Lawry
Centre for Multiple Sclerosis
Research, Munich.}
\thankstext{t2}{Supported in part by the Porticus Foundation
in the context of the International School for Clinical Medicine and
Bioinformatics.}
\affiliation{Saarland University, Helmholtz Zentrum M\"unchen and
University of Munich}
\address[A]{M. Slawski\\
Department of Computer Science\\
Saarland University \\
Saarbr\"ucken\\
Germany \\
\printead{e1}} 
\address[B]{W. zu Castell\\
Institute of Biomathematics and Biometry \\
Helmholtz Zentrum M\"unchen \\
Neuherberg\\
Germany \\
\printead{e2}}
\address[C]{G. Tutz\\
Department of Statistics\\
University of Munich \\
Munich\\
Germany\\
\printead{e3}}
\end{aug}

\received{\smonth{5} \syear{2009}}
\revised{\smonth{10} \syear{2009}}

%
\begin{abstract}
In generalized linear regression problems with an abundant number of features,
lasso-type regularization which imposes an $\ell^1$-constraint on the
regression
coefficients has become a widely established technique. Deficiencies of
the lasso in certain scenarios, notably strongly
correlated design, were unmasked when Zou and Hastie [\textit{J. Roy.
Statist. Soc. Ser. B} \textbf{67} (2005) 301--320]
introduced the elastic net. In this paper we propose to extend
the elastic net by admitting general nonnegative quadratic constraints as
a second form of regularization. The generalized ridge-type constraint will
typically make use of the known association structure of features, for
example, by
using temporal- or spatial closeness.

We study properties of the resulting ``structured elastic net''
regression estimation
procedure, including basic asymptotics and the issue of model selection
consistency. In this vein, we provide an analog to the so-called
``irrepresentable condition'' which holds for the lasso. 
Moreover, we outline algorithmic solutions for the structured elastic
net within the
generalized linear model family. The rationale and the performance of our
approach is illustrated by means of simulated and real world data, with a
focus on signal regression.
\end{abstract}



\end{frontmatter}

\section{Introduction}\label{sec1}

We consider regression problems with a linear\break predictor. Let $\mathbb
{X} =
(X_1,\ldots,X_p)^{\T}$ be a random vector of real-valued features/predictors
and let $Y$ be a random response variable taking values in a set $\mc
{Y}$. Given a realization
$\mathbf{x} = (x_1,\ldots,x_p)^{\T}$ of $\mathbb{X}$, a prediction $\wh
{y}$ for a specific
functional of the distribution of $Y|\mathbb{X} = \mathbf{x}$ is obtained
via a linear predictor
\[
f(\mathbf{x};\beta_0,\bolds{\beta}) = \beta_0 + \mathbf{x}^{\T} \bolds{\beta},
\qquad
\bolds{\beta}=(\beta_1,\ldots,\beta_p)^{\T},
\]
and a function $\zeta\dvtx \R\rightarrow\mc{Y}$ such that $\wh{y} =
\zeta(f(\mathbf{x}))$. Given an i.i.d. sample $S = \{(\mathbf{x}_i, y_i) \}_{i=1}^n$
from $(\R^p \times\mc{Y})^n$, an optimal set of coefficients $\wh
{\beta}_0$,
$\wh{\bolds{\beta}}=(\wh{\beta}_1,\ldots,\wh{\beta}_p)^{\T}$ can
be determined by
minimization of a criterion of the form
%
\begin{equation}\label{eq:remp}
(\wh{\beta}_0, \wh{\bolds{\beta}}) = \argmin\limits_{(\beta_0, \bolds{\beta})} \su L(y_i, f(\mathbf{x}_i;\beta_0, \bolds{\beta})),
\end{equation}
where $L\dvtx \mc{Y} \times\R\rightarrow  \R_0^+$ is a convex loss
function.
The loss function is chosen according to the specific prediction
problem, so that large loss represents bad fit to the observed
sample $S$. Approach \eqref{eq:remp} usually yields poor estimates
$\wh{\beta}_0, \wh{\bolds{\beta}}$ if $n$ is not one order of
magnitude larger than $p$. In particular, if $p \gg n$, approach \eqref
{eq:remp} is
not well-defined in the sense that there exist infinitely many
minimizers $\wh{\beta}_0, \wh{\bolds{\beta}}$. One way to cope with a small
$n/p$ ratio is to employ a regularizer $\Omega(\bolds{\beta})$. A
traditional approach due to
\citet{Hoe1970} minimizes the loss in equation \eqref{eq:remp}
subject to an
$\ell^2$-constraint on $\bolds{\beta}$. In the situation that $\bolds{\beta}$ is
supposed to be sparse, \citet{Tib1996} proposed, under the acronym
``lasso,'' to
work with an $\ell^1$-constraint, that is, one maximizes the loss subject
to $\Omega(\bolds{\beta}) = \Vert\bolds{\beta} \Vert_1 < s,     s > 0$.
The latter is
particularly attractive if one is interested in feature selection,
since one
obtains estimates $\wh{\beta}_j,     j \in\{1,\ldots,p\}$, which equal
exactly zero, such that feature $j$ does not contribute to prediction, for
which we say that feature $j$ is ``not selected.'' Continuous shrinkage
[\citet{FanLi2001}] and
the existence of efficient algorithms [\citet{Efron2004},
\citet{Genkin2007}]
for determining the coefficients are further virtues of the lasso. Its
limitations have recently been revealed by several
researchers. \citet{Zhou2005} pointed out that the lasso need not be unique
in the $p \gg n$ setting, where the lasso is able to select at most $n$
features [\citet{Ros2004}]. Furthermore,
Zou and Hastie stated that the lasso does not distinguish between
``irrelevant'' and ``relevant but redundant'' features. In particular, if
there is
a group of correlated features, then the lasso tends to select one arbitrary
member of the group while ignoring the remainder. The combined
regularizer of
the elastic net $\Omega(\bolds{\beta}) = \alpha\Vert\bolds{\beta}
\Vert_1
+ (1-\alpha)
\Vert\bolds{\beta} \Vert^2,     \alpha\in(0,1)$ is shown to provide
remedy in
this regard.

A second double regularizer---tailored to one-dimensional signal
regression---is employed by the fused lasso [\citet{Tib2005}], who
propagate $\Omega(\bolds{\beta}) = \alpha\Vert\bolds{\beta} \Vert_1 +
(1-\alpha) \Vert\mathbf{D} \bolds{\beta} \Vert_1$, where
\begin{eqnarray}\label{eq:fd}
\mathbf{D}\mbox{:}\qquad\hspace*{26pt}\qquad \R^p &\rightarrow&\R^{p-1},
\nonumber\\[-8pt]\\[-8pt]
(\beta_1,\ldots,\beta_p)^{\T} &\mapsto&([\beta_2 - \beta
_1],\ldots,[\beta_p - \beta_{p-1}])^{\T}\nonumber
\end{eqnarray}
is the first forward difference operator. The total variation
regularizer is
meaningful whenever there is an order relation, notably a temporal one,
among the features. The fused lasso has a property which can be
beneficial for
interpretation: it automatically clusters the features, since the sequence
$\wh{\beta}_1,\ldots,\wh{\beta}_p$ is blockwise constant.

In this paper we study a regularizer which is intermediate between the
elastic net and the fused lasso. Our regularizer
combines an $\ell^1$-constraint with a quadratic form:
%
\begin{equation}\label{eq:genetreg}
\Omega(\bolds{\beta}) = \alpha\Vert\bolds{\beta} \Vert_1 + (1-\alpha
) \bolds{\beta}^{\T}
\bolds{\Lambda} \bolds{\beta},
\end{equation}
where $\bolds{\Lambda} = (l_{jj'})_{1 \leq j,j' \leq p}$ is assumed to be
symmetric and positive semidefinite. Setting $\bolds{\Lambda} = \mathbf{I}$ yields
the elastic net. The inclusion of $\bolds{\Lambda}$ aims
at capturing the a priori association structure (if available) of the features
in more generality than the fused lasso. Therefore, expression \eqref
{eq:genetreg} will be referred to
as the structured elastic net regularizer. The structured elastic net estimator
is defined as
\begin{eqnarray}\label{eq:genet0}
&&(\wh{\beta}_0, \wh{\bolds{\beta}}) = \argmin\limits_{(\beta_0, \bolds{\beta})} \su L(y_i,
f(\mathbf{x}_i;\beta_0,\bolds{\beta})) \nonumber\\[-8pt]\\[-8pt]
&&\mbox{subject to}     \qquad \alpha\Vert\bolds{\beta} \Vert_1 +
(1-\alpha)
\bolds{\beta}^{\T} \bolds{\Lambda} \bolds{\beta} \leq s,  \qquad   \alpha\in
(0,1),s > 0,\nonumber
\end{eqnarray}
which is equivalent to the Lagrangian formulation
%
\begin{eqnarray}\label{eq:genet}
(\wh{\beta}_0, \wh{\bolds{\beta}}) = \argmin\limits_{(\beta_0, \bolds{\beta})} \su L(y_i,
f(\mathbf{x}_i;\beta_0,\bolds{\beta})) + \lambda_1 \Vert\bolds{\beta}
\Vert_1
+ \lambda_2
\bolds{\beta}^{\T} \bolds{\Lambda} \bolds{\beta}, \nonumber\\[-10pt]\\[-10pt]
\eqntext{\lambda
_1,\lambda_2 > 0.}
\end{eqnarray}
The rest of the paper is organized as follows: in Section \ref{sec2} we discuss the
choice of the matrix $\bolds{\Lambda}$, followed by an analysis of some
important properties of our proposal \eqref{eq:genet} in Section \ref{sec3}.
Section \ref{sec4}
is devoted to asymptotics and consistency questions, motivating the
introduction of the \emph{adaptive} structured elastic net. Section~\ref{sec5} presents
an algorithmic solution to compute the minimizers \eqref{eq:genet} in
the generalized linear model family. The practical performance of the
structured elastic
net is contained in Section \ref{sec6}. Section \ref{sec7} concludes with a discussion and
an outlook. All proofs can be found in the online supplement supporting this
article [\citet{Sla2009supp}].

\eject
\section{Structured features}\label{sec2}
\subsection{Motivation}\label{sec2.1}
A considerable fraction of contemporary regression problems are characterized
by a large number of features, which are either of the same order of
magnitude as the sample size or even several orders larger ($p \gg n$). Common
instances thereof are feature sets consisting of sampled signals, pixels
of an image, spatially sampled data, or gene expression intensities. Beside
high dimensionality of the feature space, these examples have in common that
the feature set can be arranged according to an a priori association
structure. If a sampled signal does not vary rapidly, the influence of nearby
sampling points on the response can be expected to be similar;
correspondingly, this applies to adjacent pixels of an image, or, more
generally, to any other form of spatially linked features. In genomics genes
can be categorized into functional groups, or one has prior knowledge
of their
functions and interactions within biochemical reaction chains, the so-called
pathways.

\begin{figure}

\includegraphics{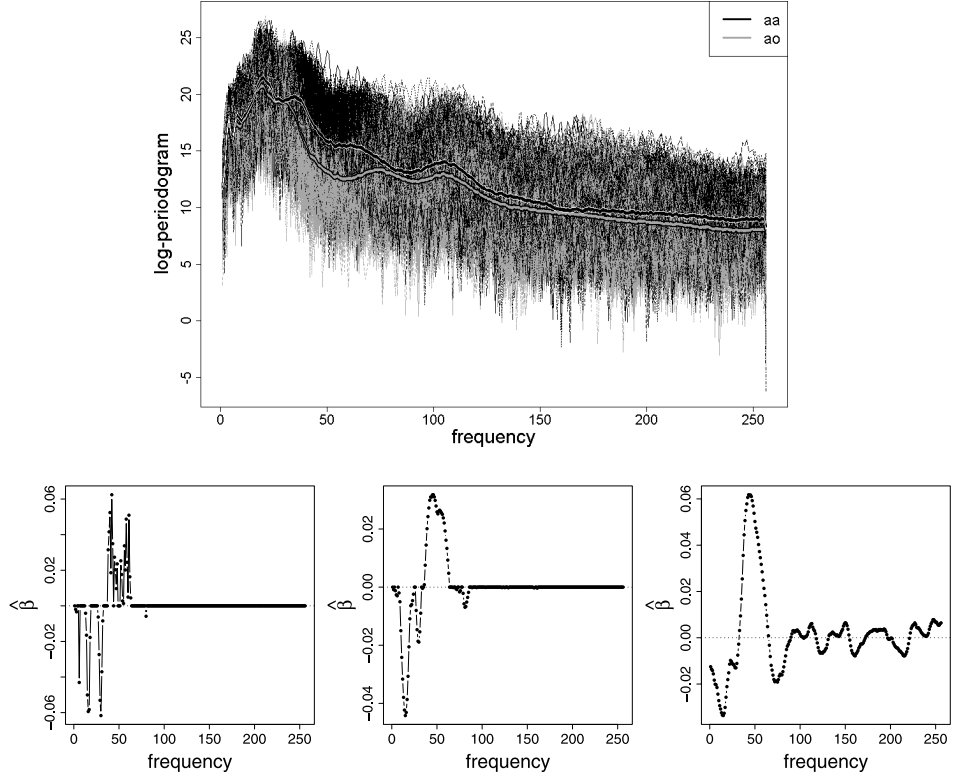}

\caption{Phoneme data [Hastie, Buja and Tibshirani (\protect\citeyear{Has1995})]. The
upper panel shows
several thousand log-periodogramms of the speech frames for the phonemes
``aa'' (as occuring in ``dark'') and ``ao'' (as occuring in ``water''). The classwise
means are given by thick lines. We use linear logistic regression to
predict the
phoneme given a log-periodogramm. The lower panel depicts the resulting
coefficients
when using the lasso (left panel), a first-order difference penalty
(right panel),
and a combination thereof, which we term ``structured elastic net''
(middle panel).}\label{fig:motivation1}
\end{figure}

\begin{figure}

\includegraphics{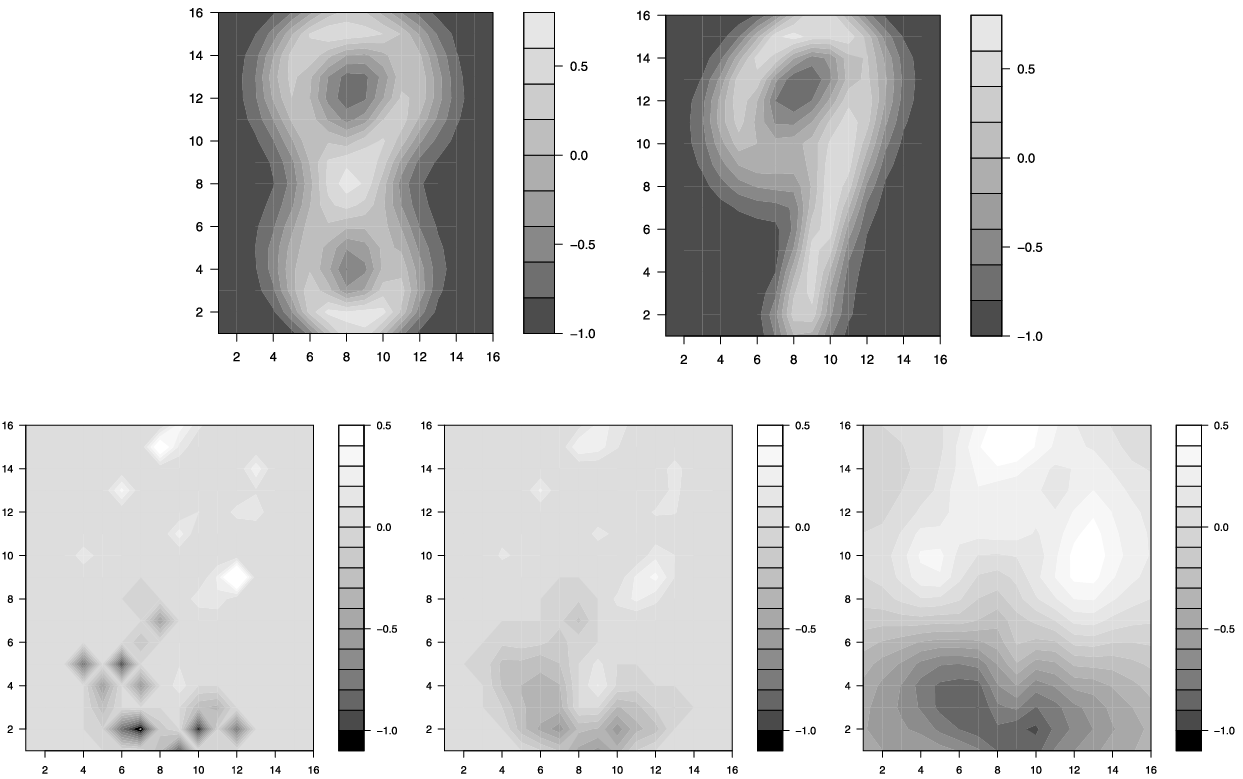}

\caption{Handwritten digit recognition data set
[Le Cun et al. (\protect\citeyear{LeCun1990})]. One
observation is given by a greyscale image
composed of 16 $\times$ 16 pixels. The upper panel shows the contour of the
pixel-wise means for the numerals ``8'' and ``9.'' We use a training set of
1500 observations of eights and nines as input for linear logistic
regression. The lower panel depicts the coefficient surfaces for the lasso
(left panel), a discrete Laplacian penalty according to the grid structure
(right panel), and a combination, the structured elastic net (middle
panel).}\label{fig:motivation2}
\end{figure}

Figures \ref{fig:motivation1} and \ref{fig:motivation2} display
two well-known examples, phoneme- and handwritten digit classification. These
examples are well apt to illustrate the idea of the structured elastic net
regularizer, since it is sensible to assume that the prediction problem
is not only characterized by smoothness
with respect to a given structure, but also by sparsity: in the phoneme
classification example, visually only the first hundred frequencies
seem to
carry information relevant to the prediction problem. A similar rationale
applies to the second example, where the arc of the numeral eight in
the lower half of
the picture is the eminent characteristic that admits a distinction
from the
numeral nine.

\subsection{Gauss--Markov random fields}\label{sec2.2}
Given a large, but structured set of features, its structure can be exploited
to cope with high dimensionality in regression estimation. The
estimands $\{
\beta_j \}_{j=1}^p$ form a finite set such that their prior dependence
structure can conveniently be described by means of a graph $\mc{G} = (V,E)$,
$V = \{\beta_1,\ldots,\beta_p \}$, $E \subset V \times V$. We
exclude loops,
that is, $(\beta_j, \beta_j) \notin E$ for all $j$. The edges may
additionally be weighted by a function $w\dvtx E
\rightarrow\R$,
$w((\beta_j, \beta_{j'})) =
w((\beta_{j'}, \beta_j))$ for all edges in $E$. We will use the notation
$\beta_j \sim\beta_{j'}$ to express that $\beta_j$ and $\beta
_{j'}$ are
connected by an edge in $\mc{G}$. The weight function can be extended
to a
function on $V \times V$ by setting $w((\beta_j,\beta_{j'})) =
w((\beta_{j'},\beta_{j})) = 0$ if $(\beta_j,\beta_{j'}) \notin
E$.\looseness=1

\begin{figure}

\includegraphics{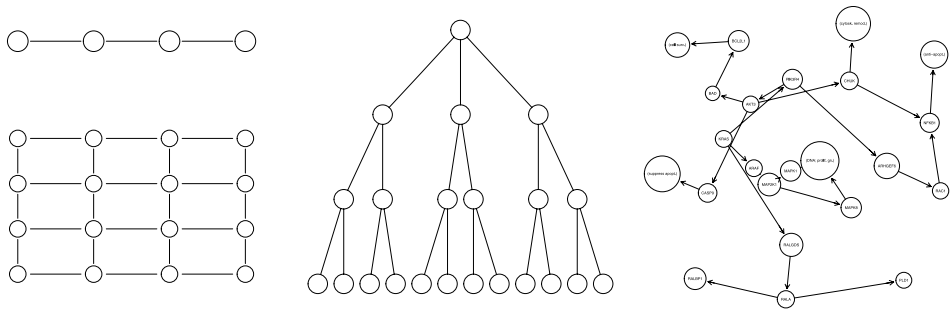}

\caption{A collection of some graphs. A path and a grid (left
panel), a rooted tree (middle panel), and an irregular graph describing a
part of the so-called MAPK signaling pathway (right panel).}\label{fig:1}
\end{figure}

The graph is interpreted in terms of the Gauss--Markov random fields
[\citet{Besag1974}; \citet{Rue2001}]. In our setup, the pairwise Markov
property reads
%
\begin{equation}\label{eq:pairwisemarkov}
\neg  \beta_j \sim\beta_{j'}  \quad \Leftrightarrow \quad \beta_j \indep
\beta_{j'} |   V
\setminus\{\beta_j, \beta_{j'} \},
\end{equation}
with $\indep$ denoting conditional independence. Property
\eqref{eq:pairwisemarkov} is conformed to the following choice for the
precision matrix
$\bolds{\Lambda} = (l_{jj'})_{1 \leq j,j' \leq p}$:
%
\begin{equation}\label{eq:prec}
l_{jj'} =
\cases{
\displaystyle\sum_{k=1}^p |w((\beta_j,\beta_k))|,  & \quad if $   j=j'$,\vspace*{2pt} \cr
-w((\beta_j,\beta_{j'})),  & \quad if $   j \neq{j'}$,
}
\end{equation}
which is singular in general.
If $\sign\{ w((\beta_j,\beta_{j'})) \} \geq0$ for all
$(\beta_j,\beta_{j'})$ in $E$, then $\bolds{\Lambda}$ as given in
equation \eqref{eq:prec} is known as the combinatorial graph Laplacian
in the spectral
graph theory [\citet{Chu1997}]. It is straightforward to verify the following
properties:
\begin{itemize}
\item
%
\begin{equation}\label{eq:energy}
\bolds{\beta}^{\T} \bolds{\Lambda} \bolds{\beta} = \sum_{\beta_j \sim
\beta_{j'}}
|w(\beta_j, \beta_{j'})|   \bigl(\beta_j - \sign \{ w((\beta_j,
\beta_{j'}))  \}   \beta_{j'}
\bigr)^2 \geq0,
\end{equation}
where the sum is over all distinct edges in $\mc{G}$, and ``distinct'' is
understood with respect to the relation $(\beta_j, \beta_{j'}) =
(\beta_{j'},
\beta_{j})$ for all $j,{j'}$.
\item
If $\mc{G}$ is connected and $\sign \{ w((\beta_j, \beta
_{j'}))  \} \geq0$ for all
$(\beta_j,\beta_{j'})$ in $E$, the null space of $\bolds{\Lambda}$ is
spanned by
the vector of ones $\mathbf{1}$.
\end{itemize}
While we have started in full generality, the choice $w((\beta_j,\beta_{j'}))
\in\{0,1\}$ for all $j,j'$ will frequently be the standard choice in
practice. In this case, the quadratic form captures local fluctuations of
$\bolds{\beta}$ w.r.t. $\mc{G}$. As a simple example, one may take $\mc
{G}$ as the
path on $p$ vertices so that expression (\ref{eq:energy}) equals the summed
squared forward differences
\setcounter{equation}{8}
\begin{equation}\label{eq:fd2}
\sum_{j=2}^p (\beta_j - \beta_{j-1})^2 = \Vert\mathbf{D} \bolds{\beta}
\Vert^2 =
\bolds{\beta}^{\T} \mathbf{D}^{\T} \mathbf{D} \bolds{\beta},
\end{equation}
where $\mathbf{D}$ is defined in equation \eqref{eq:fd}. More complex
graphical structures can be generated from simple ones using the notion
of Cartesian
products of graphs [\citet{Chu1997}, page 37]. For instance (as displayed in the left panel of Figure \ref{fig:1}), the Cartesian
product of a $p$-path and a $p'$-path equals a $p \times p'$ regular
grid, in
which case the standard choice of $\bolds{\Lambda}$ is seen to be a
discretization of the Laplacian $\Delta$ acting on functions defined on
$\R^2$. Regularizers built up from discrete differences have already
seen frequent use
in high-dimensional regression estimation. Examples comprise penalized
discriminant analysis [\citet{Has1995}] and spline smoothing
[\citet{Eil1996}].

\subsection{Connection to manifold regularization}\label{sec2.3}

As pointed out by one of the referees, regularizers of the form
(\ref{eq:energy}) are applicable to a variety of learning problems in
which data
are supposed to be generated according to a probability measure supported
on a compact, smooth manifold $M \subset\R^p$. The canonical regularization
operator acting on smooth functions on $M$ is the Laplace--Beltrami operator
$\Delta_M$ [\citet{Rosenberg1997}], which generalizes the Laplacian for
Euclidean domains. As suggested, for example, in \citet{Bel2006},
given a set of data
points in $\R^p$, a discrete proxy for a potential manifold structure
can be
obtained by computing a (possibly weighted) neighborhood graph of the points,
and, in turn, a proxy for $\Delta_M$ is obtained by a discrete
Laplacian of the
form \eqref{eq:prec} resulting from the neighborhood graph.

\begin{figure}[b]\vspace*{5pt}

\includegraphics{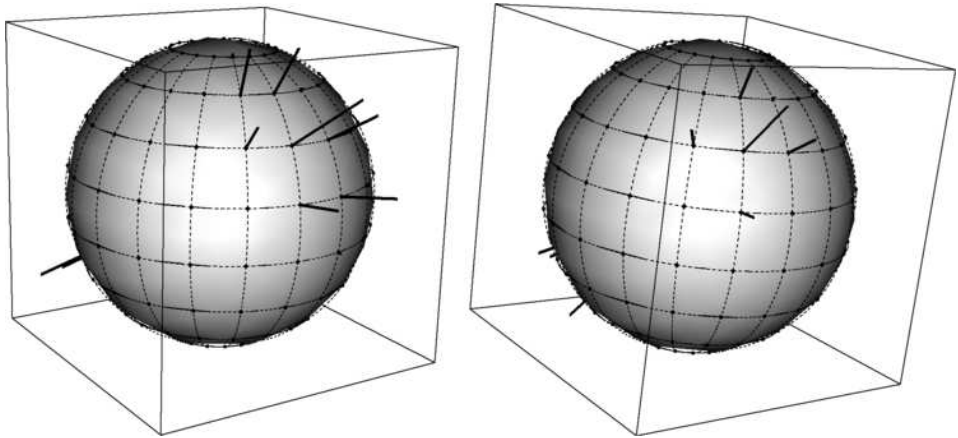}

\caption{A manifold setting suitable to our regularizer. The
black dots
represent points at which the random variables $X_j,   j=1,\ldots,p,$
are realized, and the
spikes normal to the surface indicate the size of the corresponding
$\beta_j^{\ast},   j=1,\ldots,p$. Except for the two groups
highlighted in the
left and right panel, respectively, the coefficients equal zero. The dashed
lines represent the neighborhood graph obtained by connecting each dot with
its four nearest neighbors with respect to the geodesic distance on the
sphere.}\label{fig:sphere}
\end{figure}

Relating these ideas to our framework, one might think of settings
where each
of the $\{ \mathbf{x}_i \}_{i=1}^n$ represents a collection of $p$ points sampled
on a compact, smooth manifold $M$. This is a natural extension of the
introductory examples in Section~\ref{sec2.1}, where the corresponding $M$ would be
given by an interval and a rectangle, respectively. Assuming a linear
relationship between scalar responses $\{y_i \}_{i=1}^n$ and the predictors
$\{ \mathbf{x}_i \}_{i=1}^n$, we expect the corresponding coefficient
vector to be
both sparse and smooth with respect to the manifold structure. Without going
into detail, the approach might be useful for predictors with
geographical information. The idea is illustrated in Figure \ref{fig:sphere}
where $M$ is chosen as a sphere embedded
in~$\R^3$.\

%

\section{Properties}\label{sec3}

\subsection{Bayesian and geometric interpretation}\label{sec3.1}

\begin{figure}

\includegraphics{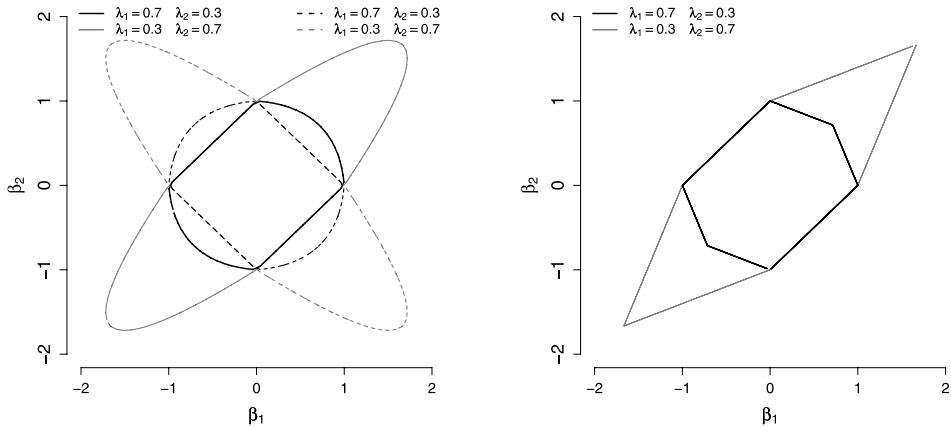}

\caption{Level sets $\{(\beta_1,\beta_2)\dvtx \lambda_1 (|\beta
_1| + |\beta_2|) +
\lambda_2 (\beta_1 - \beta_2)^2 = 1 \}$ (left panel, solid lines) and
$\{(\beta_1,\beta_2)\dvtx \lambda_1 (|\beta_1| + |\beta_2|) +
\lambda_2 (\beta_1 + \beta_2)^2 = 1 \}$ (left panel, dashed lines)
of the
structured elastic net regularizer and
$\{(\beta_1,\beta_2)\dvtx \lambda_1 (|\beta_1| + |\beta_2|) +
\lambda_2 |\beta_1 - \beta_2| = 1 \}$ of the fused lasso.}
\label{fig:2}
\end{figure}

In the setup of Section \ref{sec1}, consider the regularizer
\[
\Omega(\bolds{\beta}) = \lambda_1 \Vert\bolds{\beta} \Vert_1 +
\lambda_2
\bolds{\beta}^{\T}
\bolds{\Lambda} \bolds{\beta}, \qquad    \lambda_1, \lambda_2 > 0.
\]
It has a nice Bayesian interpretation when the loss function $L$ is of the
form
%
\begin{equation}\label{eq:glmloss}
L(y,f(\mathbf{x};\beta_0,\bolds{\beta})) = \phi^{-1} \bigl( b(f(\mathbf{x})) - y
f(\mathbf{x})\bigr) + c(y,\phi),
\end{equation}
that is, the loss function equals the negative log-likelihood of a generalized
linear model in canonical parametrization, which will primarily be
studied in this paper. Models of this
class are characterized by [cf. \citet{Mcc1989}]
\begin{eqnarray}\label{eq:glms}
Y | \mathbb{X} &=& \mathbf{x}   \sim  \mbox{simple exponential
family}, \nonumber\\
\wh{y} &=& \E[Y|\mathbb{X} = \mathbf{x}] = \mu= \frac{d}{df} b(f(\mathbf{x})), \\
\var[Y|\mathbb{X} = \mathbf{x}] &=& \phi\frac{d^2}{df^2} b(f(\mathbf{x})).\nonumber
\end{eqnarray}
The form \eqref{eq:glmloss} is versatile, including classical linear
regression with Gaussian errors, logistic regression for
classification, and
Poisson regression for count data. Given a loss from the class
\eqref{eq:glmloss}, the regularizer $\Omega(\bolds{\beta})$ can be
interpreted as
the combined Laplace (double exponential)-Gaussian prior $p(\bolds{\beta}) \propto
\exp(-\Omega(\bolds{\beta}))$, for which the structured elastic net estimator
\eqref{eq:genet}, provided $p(\beta_0) \propto1$, is the maximum
posterior (MAP) estimator given the sample
$S$.
It is instructive to consider two predictors, that is, $\bolds{\beta} =
(\beta_1,\beta_2)^{\T}$. Figure \ref{fig:2} gives a geometric interpretation
for the basic choices
\[
\bolds{\Lambda} =  \pmatrix{ 1 & -1 \cr -1 & 1}  \quad\mbox{and}\quad
\bolds{\Lambda} =  \pmatrix{ 1 & 1 \cr 1 & 1 },
\]
corresponding to positive- and negative prior correlation, respectively.

The contour lines of the structured elastic net penalty contain
elements of a
diamond and an ellipsoid. The higher $\lambda_2$ in relation to
$\lambda_1$,
the ellipsoidal part becomes more narrower and more stretched. The sign
of the
off-diagonal element of $\bolds{\Lambda}$ determines the orientation of the
ellipsoidal part.

\subsection{A grouping property}\label{sec3.2}
For the elastic net, \citet{Zhou2005} provided an upper bound on the absolute
distances $|\wh{\beta}{}_{j}^{\mathrm{elastic\ net}} - \wh{\beta
}{}_{j'}^{\mathrm{elastic\ net}}|,   j,j'=1,\ldots,p$, in terms of the sample correlations, to which
Zou and Hastie referred to as ``grouping property.'' We provide similar
bounds here. For what follows, let $S$ be a sample as in Section \ref{sec1}. We
introduce a
design matrix $\mathbf{X} = \mathop{(x_{ij})_{1 \leq i \leq n }}\limits_{\qquad 1 \leq
j \leq p}$ and denote by $\mathbf{X}_j = (x_{1j},\ldots,x_{nj})^{\T}$ the
realizations of predictor $j$ in $S$, and the response vector is
defined by
$\mathbf{y}=(y_1,\ldots,y_n)^{\T}$. For the remainder of this section,
we assume
that the responses are centered and that the predictors are centered and
standardized to unit Euclidean length w.r.t. the sample $S$, that is,
%
\begin{equation}
\sum_{i=1}^n y_i = \sum_{i=1}^n x_{ij} = 0, \qquad \sum_{i=1}^n
x_{ij}^2 = 1,  \qquad   j=1,\ldots,p.
\end{equation}

\begin{prop}\label{prop1}
Letting $p = 2$, let the loss function be of the form\vspace*{1pt}
\eqref{eq:glmloss}, let $\rho= \mathbf{X}_1^{\T} \mathbf{X}_2$ denote the sample
correlation of $\mathbf{X}_1$ and $\mathbf{X}_2$, and let $\bolds{\Lambda} =
\frac{1}{2}  { 1 \ s \choose s \ 1}$, $s \in\{-1,1\}$. If
$-s \wh{\beta}_1 \wh{\beta}_2 > 0$, then
\[
|\wh{\beta}_1 + s \wh{\beta}_1| \leq\frac{1}{2 \lambda_2} \sqrt
{2 (1 +
s \rho)} \Vert\mathbf{y} \Vert.
\]
\end{prop}

In particular, in the setting of Proposition \ref{prop1}, we have the implication that
if $\mathbf{X}_1 = -s \mathbf{X}_2$, then $\wh{\beta}_1 = -s \wh{\beta
}_2$.

\subsection{Decorrelation}\label{sec3.3}

Let us now consider the important special case
\[
L(y, f(\mathbf{x}; \bolds{\beta})) = (y - \mathbf{x}^{\T} \bolds{\beta})^2,
\]
which corresponds to classical linear regression. The constant term
$\beta_0$ is
omitted, since we work with centered data. The structured elastic net
estimator can then be written as
\begin{eqnarray}\label{eq:genetls}
\qquad\wh{\bolds{\beta}} &=& \argmin\limits_{\bolds{\beta}} - 2 \mathbf{y}^{\T} \mathbf{X}
\bolds{\beta} +
\bolds{\beta}^{\T} [\mathbf{C} + \lambda_2 \bolds{\Lambda}] \bolds{\beta} +
\lambda_1 \Vert\bolds{\beta} \Vert_1,\qquad
\mathbf{C} = \mathbf{X}^{\T} \mathbf{X}, \nonumber\\[-8pt]\\[-8pt]
&=& \argmin\limits_{\bolds{\beta}} - 2 \mathbf{y}^{\T} \mathbf{X} \bolds{\beta} + \bolds{\beta}^{\T}
\wt{\mathbf{C}} \bolds{\beta} + \lambda_1 \Vert\bolds{\beta}
\Vert_1,\qquad
\wt{\mathbf{C}} =
\mathbf{X}^{\T} \mathbf{X} + \lambda_2 \bolds{\Lambda}.\nonumber
\end{eqnarray}
Note that for standardized predictors, $\mathbf{C}$ equals the matrix of sample
correlations $\rho_{jj'} = \mathbf{X}_j^{\T} \mathbf{X}_{j'},
j,j'=1,\ldots,p$. With\vspace*{1pt} a large number of predictors or elements
$\rho_{jj'}$ with large $|\rho_{jj'}|$, $\mathbf{C}$ is known to yield
severely unstable
ordinary least squares (ols)\vspace*{1pt} estimates $\wh{\beta}_j^{\mathrm{ols}},
j=1,\ldots,p$. If the two underlying random variables $X_j$ and
$X_{j'}$ are
highly positively correlated, this will likely translate to high sample
correlations of $\mathbf{X}_j$ and $\mathbf{X}_{j'}$, which in turn yield a strongly
negative correlation between $\wh{\beta}_j^{\mathrm{ols}}$ and
$\wh{\beta}_{j'}^{\mathrm{ols}}$ and, as a consequence, high variances
$\var[\wh{\beta}_j^{\mathrm{ols}}]$ and $\var[\wh{\beta
}_{j'}^{\mathrm{ols}}]$.
In the prevalence of high correlations, performance of the lasso may
degrade as well. For example, \citet{Don2006} showed that the lower the\vspace*{-2pt}
\textit{mutual coherence} $\mathop{\max_{j,j' }}\limits_{\hspace*{18pt} j \neq j'} |\rho
_{j,j'}|$, the more stable is lasso estimation.
The modified matrix $\wt{\mathbf{C}}$ 
can be written as $\wt{\mathbf{C}} = \mathbf{V}_{\bolds{\Lambda}}^{1/2}
\mathbf{R}_{\bolds{\Lambda}} \mathbf{V}_{\bolds{\Lambda}}^{1/2}$, $\mathbf{V}_{\bolds{\Lambda}} =
\diag(1 + \lambda_2 \sum_{k=1}^p |l_{1k}|,\ldots, 1 + \lambda_2
\sum_{k=1}^p
|l_{pk}|)$, and the modified correlation matrix $\mathbf{R}_{\bolds{\Lambda}}$ has entries
\[
\rho_{\bolds{\Lambda},jj'} = \frac{\rho_{jj'} + \lambda_2
l_{jj'}}{\sqrt{1 +
\sum_{k=1}^p |l_{jk}|}\sqrt{1 + \sum_{k=1}^p |l_{j'k}|}},\qquad
j,j'=1,\ldots,p.
\]
In light of Section \ref{sec2}, the entries of $\mathbf{R}_{\bolds{\Lambda}}$
combine sample- and prior
correlations. Decorrelation occurs if $\rho_{jj'} \approx
-\lambda_2 l_{jj'}$.

\section{Consistency}\label{sec4}

The asymptotic analysis presented in this section closely follows the ideas
of \citet{KnightFu2000} and \citet{Zhou2006}. Both have studied
asymptotics for
the lasso in linear regression for a fixed number of predictors under
conditions ensuring $\sqrt{n}$-consistency and asymptotic normality of
the ordinary least squares estimator.
%
\citet{KnightFu2000} proved that the lasso estimator $\wh{\bolds{\beta}}{}^{\mathrm{lasso}}$ is $\sqrt{n}$-consistent for
the true coefficient vector $\bolds{\beta}^{\ast}$ provided $\lambda
_1^n = O(\sqrt{n})$.
%
\citet{Zhou2006} has shown that while this choice of $\lambda_1^n$
provides the optimal rate for estimation, it leads to inconsistent feature
selection. Define the active set as $A = \{j\dvtx \beta_j^{\ast} \neq0 \}
$ and
$A^c = \{1,\ldots,p \} \setminus A$ and let $\delta$ be an estimation
procedure producing an estimate $\wh{\bolds{\beta}}{}^{\delta}$. Then
$\delta$ is
said to be selection consistent if
\begin{eqnarray*}
\lim_{\ninf} \p(\wh{{\beta}}_{j,n}^{\delta} \neq0) &=& 1\qquad
\mbox{for }   j \in A,\\
\lim_{\ninf} \p(\wh{{\beta}}_{j,n}^{\delta} = 0) &=& 1  \qquad\mbox{for }   j \in A^c,
\end{eqnarray*}
where here and in the following, the sub- or superscript $n$ indicates
that the corresponding quantity depends on the sample size $n$.
Moreover, \citet{Zhou2006} and \citet{ZhaoYu2006} have shown that if
$\lambda_1^n = o(n)$ and $\lambda_1^n/\sqrt{n} \rightarrow\infty$,
the lasso has to satisfy a nontrivial condition, the so-called
``irrepresentable condition,'' to be selection consistent. \citet{Zhou2006}
proposed the adaptive lasso, a two-step estimation procedure, to fix this
deficiency. In the following, these results will be adapted to the
presence of
a second quadratic penalty term.

\begin{theorem}\label{thm1}
Define
\[
\wh{\beta}_n = \argmin\limits_{\bolds{\beta}} \Vert\mathbf{y}_n - \mathbf{X}_n
\bolds{\beta} \Vert^2 +
\lambda_1^n \Vert\bolds{\beta} \Vert_1 + \lambda_2^n \bolds{\beta}^{\T}
\bolds{\Lambda} \bolds{\beta}.
\]
Assume that $\lambda_1^n/\sqrt{n} \rightarrow\lambda_1^0 \geq0$ and
$\lambda_2^n/\sqrt{n} \rightarrow\lambda_2^0 \geq0$. Consider the
random function
\begin{eqnarray*}
V(\mathbf{u}) &=& - 2 \mathbf{u}^{\T} \mathbf{w} + \mathbf{u}^{\T} \mathbf{C} \mathbf{u}
\\
&&{} + \lambda_1^0 \sum_{j=1}^p u_j \sign(\beta_j^{\ast}) I(\beta
_j^{\ast} \neq0) +
|u_j| I(\beta_j^{\ast} = 0) \\
&&{} + 2 \lambda_2^0 \mathbf{u}^{\T} \bolds{\Lambda} \bolds{\beta}^{\ast},
  \qquad   \mathbf{w} \sim  N(\mathbf{0}, \sigma^2 \mathbf{C}).
\end{eqnarray*}
Then, under conditions \textup{(C.1)--(C.3)} in the online supplement, $\sqrt
{n}(\wh{\bolds{\beta}}_n - \bolds{\beta}^{\ast}) \stackrel{D}{\rightarrow} \argmin
V(\mathbf{u})$.
\end{theorem}

Theorem \ref{thm1} is analogous to Theorem 2 in \citet{KnightFu2000} and establishes
$\sqrt{n}$-consistency of $\wh{\bolds{\beta}}_n$, provided $\lambda
_1^n$ and
$\lambda_2^n$ are $O(\sqrt{n})$. Theorem \ref{thm1} admits a straightforward
extension to the
class of generalized linear models [cf. equation~\eqref{eq:glms}]. Let
the true model
be defined by
\[
\E[Y|\mathbb{X} = \mathbf{x}] = b'(f(\mathbf{x};\bolds{\beta}^{\ast})),
\qquad
f(\mathbf{x}) = \mathbf{x}^{\T} \bolds{\beta}^{\ast}.
\]
For the sake of a clearer presentation, we assume that $\beta_0^{\ast
} = 0$. We
study the estimator
%
\begin{equation}\label{eq:glmestcons}
\qquad\wh{\bolds{\beta}}_n = \argmin\limits_{\bolds{\beta}} 2 \phi^{-1} \su
b(f(\mathbf{x}_i;\bolds{\beta})) - y_i f(\mathbf{x}_i;\bolds{\beta}) + \lambda_1^n
\Vert\bolds{\beta} \Vert_1 + \lambda_2^n \bolds{\beta}^{\T} \bolds{\Lambda}
\bolds{\beta}.
\end{equation}
%

\begin{theorem}\label{thm2}
For the estimator \eqref{eq:glmestcons}, let $\lambda_1^n/\sqrt{n} \rightarrow\lambda_1^0
\geq0$ and $\lambda_2^n/\sqrt{n} \rightarrow\lambda_2^0 \geq0$. Consider
the random function
\begin{eqnarray*}
W(\mathbf{u}) &=& - 2 \mathbf{u}^{\T} \mathbf{w} + \mathbf{u}^{\T} \mc{I} \mathbf{u}
\\
&&{} + \lambda_1^0 \sum_{j=1}^p u_j \sign(\beta_j^{\ast}) I(\beta
_j^{\ast} \neq0) +
|u_j| I(\beta_j^{\ast} = 0) \\
&&{} + 2 \lambda_2^0 \mathbf{u}^{\T} \bolds{\Lambda} \bolds{\beta}^{\ast},
  \qquad   \mathbf{w}
\sim  N(\mathbf{0}, \mc{I}).
\end{eqnarray*}
Then under conditions \textup{(G.1)} and \textup{(G.2)} in the online supplement, $\sqrt
{n}(\wh{\bolds{\beta}}_n - \bolds{\beta}^{\ast}) \stackrel{D}{\rightarrow} \argmin
W(\mathbf{u})$.
\end{theorem}

Now let us turn to the question of selection consistency. In the setup of
Theorem 1, if $\lambda_1^n$ and $\lambda_2^n$ both are $O(\sqrt
{n})$, then,
using arguments similar to those in \citet{KnightFu2000} and \citet{Zhou2006},
$\wh{\bolds{\beta}}_n$ is shown not to be selection consistent. 
Selection consistency can be achieved if one lets $\lambda_1^n,
\lambda_2^n$ grow more strongly and
if the quantities $\mathbf{C}$, $\bolds{\Lambda},$ and $\bolds{\beta}^{\ast
}$ jointly fulfill a nontrivial
condition, which can be seen as analog to the \emph{irrepresentable condition}
of the lasso [\citet{Zhou2006}, \citet{ZhaoYu2006}]. 

\begin{theorem}\label{thm3}
In the situation of Theorem \ref{thm1}, let $\lambda_1^n/n
\rightarrow
0$, $\lambda_1^n/\sqrt{n} \rightarrow\infty$, $\lambda_2^n/\lambda_1^n
\rightarrow R,   0 < R < \infty$ and consider the partitioning scheme
%
\begin{eqnarray}\label{eq:partioning}
\bolds{\beta}^{\ast} &=&  \pmatrix{
\bolds{\beta}_A^{\ast} \vspace*{2pt}\cr
\bolds{\beta}_{A^c}^{\ast}
},\qquad
\mathbf{C} =  \pmatrix{\mathbf{C}_A & \mathbf{C}_{AA^c} \vspace*{2pt}\cr \mathbf{C}_{A^c A} & \mathbf{C}_{A^c}
}
\quad   \mbox{and}\nonumber\\[-8pt]\\[-8pt]
\bolds{\Lambda} &=&  \pmatrix{ \bolds{\Lambda}_A & \bolds{\Lambda}_{AA^c} \vspace*{2pt}\cr \bolds{\Lambda}_{A^c A}
& \bolds{\Lambda}_{A^c} },\nonumber
\end{eqnarray}
so that here and in the following, the subscripts $A$ and $A^c$ refer
to active and inactive set, respectively.
Then, if selection consistency holds, the following condition must be
fulfilled: there exists a sign vector $\mathbf{s}_A$
such that
\[
|{-}\mathbf{C}_{A^c A} \mathbf{C}_A^{-1}(\mathbf{s}_A + 2 R \bolds{\Lambda}_A
\bolds{\beta}_A^{\ast}) + 2 R \bolds{\Lambda}_{A^c A} \bolds{\beta}_A^{\ast}| \leq\mathbf{1},
\]
where the inequality is interpreted componentwise.
\end{theorem}

While this condition is interesting from a theoretical point of view,
it is
impossible to check in practice, since $\bolds{\beta}_A^{\ast}$ is
unknown.

Selection consistency can be achieved by a two-step estimation strategy
introduced in \citet{Zhou2006} under the name adaptive lasso, which replaces
$\ell^1$-regularization uniform in $\beta_j,  j=1,\ldots,p,$ by a weighted
variant $J(\bolds{\beta}) = \sum_{j=1}^p \omega_j |\beta_j|$, where
the weights
$\{ \omega_j \}_{j=1}^p$ are determined adaptively as a function of an ``initial
estimator'' $\wh{\bolds{\beta}}{}^{\mathrm{init}}$:
%
\begin{equation}\label{eq:adaweights}
\omega_j = |\wh{\beta}{}_j^{\mathrm{init}}|^{-\gamma},  \qquad   \gamma>
0,     j=1,\ldots,p.
\end{equation}
In terms of selection consistency, this strategy turns out to be
favorable for
our proposal, too.

\begin{theorem}\label{thm4}
In the situation of Theorem \ref{thm1}, define
\[
\wh{\bolds{\beta}}{}_n^{\mathrm{adaptive}} = \argmin\limits_{\bolds{\beta}} \Vert
\mathbf{y}_n - \mathbf{X}_n \bolds{\beta} \Vert^2 + \lambda_1^n \sum
_{j=1}^p \omega_j |\beta
_j| +
\lambda_2^n \bolds{\beta}^{\T} \bolds{\Lambda} \bolds{\beta},
\]
where the weights are as in equation \eqref{eq:adaweights}, and
suppose that the initial
estimator satisfies
\[
r_n (\wh{\bolds{\beta}}_n - \bolds{\beta}^{\ast}) = O_{\p}(1),\qquad
r_n \rightarrow\infty  \mbox{ as }   \ninf.
\]
Furthermore, suppose that
\[
r_n^{\gamma} \lambda_1^n n^{-1/2} \rightarrow\infty,\qquad
\lambda_1^n n^{-1/2} \rightarrow0,\qquad
\lambda_2^n n^{-1/2} \rightarrow\lambda_2^0 \geq0
\]
as $\ninf$. Then:
\begin{enumerate}[(2)]
\item[(1)]  $   \sqrt{n}(\wh{\bolds{\beta}}{}_{A,n}^{\mathrm{adaptive}} - \bolds{\beta}_A^{\ast})
\stackrel{D}{\rightarrow} N(-\lambda_2^0 \mathbf{C}_A^{-1} \bolds{\Lambda}_A \bolds{\beta}_A^{\ast},
\mathbf{C}_A^{-1})$,
\item[(2)] $    \lim_{\ninf} \p(\wh{\bolds{\beta}}{}_{A^c,n}^{\mathrm{adaptive}} = \mathbf{0}) = 1$.
\end{enumerate}
\end{theorem}

Theorem \ref{thm4} implies that the adaptive structured elastic net
$\wh{\bolds{\beta}}{}^{\mathrm{adaptive}}$ is an oracle estimation procedure
[\citet{FanLi2001}] if the bias term in (1) vanishes, which is the
case if
$\bolds{\beta}_A^{\ast}$ resides in the null space of $\bolds{\Lambda}_A$. Interestingly,
if $\bolds{\Lambda}$ equals the combinatorial graph Laplacian (cf.
Section \ref{sec2.2}), this
happens if and only if $\bolds{\beta}_A^{\ast}$ has constant entries
and $A$ specifies a connected component in the underlying
graph.\looseness=1

Concerning the choice of the initial estimator, the ridge estimator has
worked well
for us in practice, provided the ridge parameter is chosen
appropriately. While
$\gamma$ may be treated as a tuning parameter, we have set $\gamma$
equal to 1
in all our data analyses. Last, we remark that while Theorem \ref{thm4} applies
to linear regression, it can be extended
to hold for generalized linear models, similarly as we have extended
Theorem \ref{thm1} to Theorem \ref{thm2}.

\section{Computation}\label{sec5}

This section discusses aspects concerning computation and model
selection for
the structured elastic net estimator when the loss function is the negative
log-likelihood of a generalized linear model \eqref{eq:glmloss}.

\subsection{Data augmentation}\label{sec5.1}

From the discussions in Section \ref{sec3.3}, it follows that the
structured elastic net for squared loss, assuming centered data, can be
recast as the lasso on augmented data
\[
\wt{\mathbf{X}} =  \pmatrix{
\mathbf{X} \vspace*{3pt}\cr
\lambda_2^{1/2} \mathbf{Q}
}_{(n+p) \times p}, \qquad \wt{\mathbf{y}} =  \pmatrix{
\mathbf{y} \vspace*{2pt}\cr
\mathbf{0}
}_{(n+p) \times1}, \qquad \bolds{\Lambda} =
\mathbf{Q}^{\T} \mathbf{Q},
\]
and, hence, algorithms available for computing the lasso, notably LARS
[\citet{Efron2004}], may be applied, which computes for fixed $\lambda
_2$ and
varying $\lambda_1$ the piecewise linear solution path
$\wh{\bolds{\beta}}(\lambda_1;\lambda_2)$. This approach is parallel
to that
proposed by Zou and Hastie (\citeyear{Zhou2005}) for the elastic net. In addition, the
augmented data representation is helpful when addressing uniqueness of the
structured elastic net in the $p \gg n$ setting: if $\operatorname{rank}(\mathbf{X}) +
\operatorname{rank}(\lambda_2^{1/2} \mathbf{Q}) \geq p$ and the rows of $\mathbf{X}$ combined
with the rows of $\lambda_2^{1/2} \mathbf{Q}$ form a linearly independent set,
$\wt{\mathbf{C}}$ as defined in equation \eqref{eq:genetls} is of full
rank and, hence, the structured elastic net is
unique. Moreover, this shows that even for $p \gg n$, in principle, all
features can be selected.

In order to fit arbitrary regularized generalized linear models, the augmented
data representation has to be modified. Without regularization,
estimators in generalized linear models are obtained by
iteratively
computing weighted least squares estimators:
\begin{eqnarray}\label{eq:iwls}
\pmatrix{
\wh{\beta}{}_0^{(k+1)} \vspace*{3pt}\cr
\wh{\bolds{\beta}}{}^{(k+1)}
} &=&  \bigl([\matrix{\mathbf{1}  & \mathbf{X}}]^{\T}
\mathbf{W}\ik[\matrix{\mathbf{1}& \mathbf{X}}]  \bigr)^{-1} [\matrix{\mathbf{1}  & \mathbf{X}}]^{\T} \mathbf{W} \ik\mathbf{z}
\ik,\nonumber\\
\mathbf{z} \ik&=& \mathbf{f} \ik+ \bigl[\mathbf{W} \ik\bigr]^{-1} \bigl(\mathbf{y} - \bolds{\mu}\ik\bigr),\nonumber\\
\mathbf{f} \ik&=& \bigl(f_1 \ik,\ldots,f_n \ik\bigr)^{\T}, \qquad  f_i \ik= \wh
{\beta}{}_0 \ik+
\mathbf{x}_i^{\T} \wh{\bolds{\beta}}{} \ik,     i=1,\ldots,n, \\
\bolds{\mu} \ik&=& \bigl(\mu_1 \ik,\ldots,\mu_n \ik\bigr)^{\T}, \qquad \mu
_i \ik= b'\bigl(f_i \ik\bigr),
i=1,\ldots,n, \nonumber\\
\mathbf{W} \ik&= &\diag\bigl(w_1 \ik,\ldots, w_n \ik\bigr),\qquad   w_i \ik= \phi
^{-1} b''\bigl(f_i \ik\bigr),     i=1,\ldots,n.\nonumber
\end{eqnarray}
Note that the design matrix additionally includes a constant term $\mathbf{1}$.
Turning back to the structured elastic net, an adaptation of the
augmented data
approach iteratively determines
\[
\pmatrix{
\wh{\beta}{}_0^{(k+1)}\vspace*{3pt} \cr
\wh{\bolds{\beta}}{}^{(k+1)}
} = \argmin\limits_{(\beta_0, \bolds{\beta})} \sum_{i=1}^{n+p} \wt
{w}_i \ik \left(\wt{z}_i \ik- \wt{\mathbf{x}}_i^{\T}  \pmatrix{
\beta_0 \vspace*{2pt}\cr
\bolds{\beta}
}\right)^2 + \lambda_1 \Vert\bolds{\beta} \Vert_1,
\]
with
\begin{eqnarray*}
\wt{w}_i \ik&=& w_i \ik, \qquad    i=1,\ldots,n,   \qquad\mbox{as in
equation \eqref{eq:iwls}},\\
 \wt{w}_i \ik &=& 1,  \qquad   i=(n+1),\ldots,(n+p),\\
\wt{z}_i \ik&=& z_i \ik  ,\qquad  i=1,\ldots,n,   \qquad\mbox{as in
equation \eqref{eq:iwls}}, \\
\wt{z}_i \ik&=& 0, \qquad    i=(n+1),\ldots
,(n+p),\\
\wt{\mathbf{x}}_i& =& \pmatrix{1 &  \mathbf{x}_i^{\T}}^{\T},   \qquad
i=1,\ldots,n,\\
\wt{\mathbf{x}}_i &=& \pmatrix{0  & \sqrt{\lambda_2}
\mathbf{q}_i^{\T}}^{\T},\qquad
  i=(n+1),\ldots,(n+p),
\end{eqnarray*}
with $\mathbf{q}_i^{\T}$ denoting the $i$th row of $\mathbf{Q}$.

Alternatives to augmented data representation include cyclical
coordinate descent
in the spirit of \citet{Friedman2007} and a direct modification of Goeman's
algorithm [\citet{Goeman2007a}]. Descriptions can be a found in the full
technical report underlying this article [\citet{Sla2009}, available
online].\looseness=1

\section{Data analysis}\label{sec6}

\subsection{One-dimensional signal regression}\label{sec6.1}

In one-dimensional signal regression, as described, for example, in
\citet{Fra1993}, one aims at the
prediction of a response given a sampled signal $\mathbf{x}^{\T} =
(x(t))_{t=1}^T$, where the indices $t=1,\ldots,T$, refer to different
ordered sampling
points. For a sample $S = \{ ( \{ x_1 (t) \}_{t=1}^T, y_1), \ldots,
(\{ x_n
(t) \}_{t=1}^T, y_n) \}$ of pairs consisting of sampled signals and responses,
we consider prediction models of the form
\[
\wh{y}_i = \zeta \Biggl( \wh{\beta}_0 + \sum_{t=1}^T x_i(t) \wh
{\beta}(t)
 \Biggr),\qquad   i=1,\ldots,n.
\]

\subsubsection{Simulation study}\label{sec6.1.1}

\begin{figure}[b]

\includegraphics{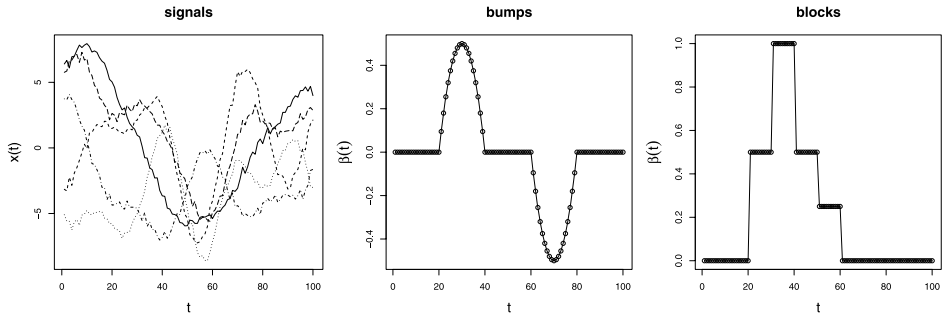}

\caption{The setting of the
simulation study. A collection of five signals (left
panel), the coefficient functions for ``bump''---(middle panel) and
``block'' setting (right panel), respectively.}
\label{fig:simulationsetupsignals}
\end{figure}

Similarly to \citet{Tutz2009}, we simulate signals $x(t),
t=1,\ldots,T$,     $T=100$, according to
\begin{eqnarray*}
\{ x(t) \}   &\sim & \sum_{k=1}^5 b_{k} \sin\bigl(t \pi(5 - b_{k})/50 -
m_{k}\bigr) + \tau(t), \\
\{ b_{k} \}   &\sim&  U(0,5), \qquad \{ m_{k} \}   \sim  U(0, 2 \pi ),
\qquad \{ \tau(t) \} \sim N(0, 0.25),
\end{eqnarray*}
with $U(a,b)$ denoting the uniform distribution on the interval
$(a,b)$. For the coefficient function $\beta^{\ast}(t),   t=1,\ldots
,T$, we examine
two cases. In the first case, referred to as the ``bump setting,'' we use
\[
\beta^{\ast}(t) =
\cases{
- \{ (30 - t)^2 + 100  \}/200,  &\quad $t=21,\ldots,39$, \cr
 \{ (70 - t)^2 - 100  \}/200,  &\quad $t=61,\ldots,80$, \cr
0,  & \quad otherwise.
}
\]
In the second case, referred to as the ``block setting,''
\[
\bolds{\beta}^{\ast} = (\underbrace{0,\ldots,0}_{20
\ \mathrm{times}},\underbrace{0.5,\ldots,0.5}_{10
\ \mathrm{times}},\underbrace{1,\ldots,1}_{10
\ \mathrm{times}},\underbrace{0.5,\ldots,0.5}_{10
\ \mathrm{times}},\underbrace{0.25,\ldots,0.25}_{10
\ \mathrm{times}},\underbrace{0,\ldots,0}_{40 \ \mathrm{times}})^{\T}.
\]
The form of the signals and coefficient functions are displayed in Figure
\ref{fig:simulationsetupsignals}.

For both settings, data are simulated according to
\[
y = \sum_{t=1}^T x(t) \beta^{\ast}(t) + \epsilon,\qquad
\epsilon \sim  N(0,5).
\]
For each out of 50 iterations, we simulate $i=1,\ldots,500$ i.i.d.
realizations and divide them
into three parts: a training set of size 200, a validation set of size
100, and
a test set of size 200. Hyperparameters of the methods listed below are
optimized by means of the validation set. As performance measures, we compute
the absolute distance $L^1(\wh{\bolds{\beta}},\bolds{\beta}) = \Vert\wh
{\bolds{\beta}} - \bolds{\beta} \Vert_1$ of true- and estimated
coefficients and the mean
squared prediction error on the test
set. For methods with built-in feature selection, we additionally
evaluate the goodness
of selection in terms of sensitivity and specificity. 
%
For each of the two setups, the simulation is repeated $50$ times. The
following methods are compared: ridge regression, generalized ridge
regression with a first difference penalty,
P-splines according to \citet{Eilers1999}, lasso, fused lasso, elastic
net, structured elastic net with a first difference penalty,
adaptive structured elastic net, where the weights $\{ \omega(t) \}$
are chosen according to the ridge estimator of the same iteration as
$\omega(t) = 1/|\wh{\beta}^{\mathrm{ridge}}(t)|$.

Performance measures are averaged over 50 iterations and displayed in
Table~\ref{tab:bumpsetting} (bump setting) and Table \ref{tab:blocksetting} (block
setting), respectively.

\begin{table}[b]
\caption{Results for the bump setting, averaged over 50
simulations}\label{tab:bumpsetting}
\begin{tabular*}{\textwidth}{@{\extracolsep{\fill}}lcccc@{}}
\hline
\textbf{Method} & $\bolds{L^1(\wh{\bolds{\beta}}, \bolds{\beta}^{\ast})}$ & \textbf{PE} &
\textbf{Sensitivity} & \textbf{Specificity} \\
\hline
Ridge & 0.249 & 5.35 & & \\
& ($5.9 \times10^{-4}$) & (0.078) & & \\[3pt]
G.ridge & 0.238 & 5.32 & & \\
& ($9.9 \times10^{-4}$) & (0.076) & & \\[3pt]
P-spline & 0.241 & 5.30 & & \\
& ($16.0 \times10^{-4}$) & (0.077) & & \\[3pt]
Lasso & 0.271 & 5.72 & 0.62 & 0.65 \\
& ($23.9 \times10^{-4}$) & (0.079) & ($8.9 \times10^{-3}$)& (0.016) \\[3pt]
Fused lasso & 0.235 & 5.30 & 0.96 & 0.51 \\
& ($7.2 \times10^{-4}$) & (0.075) & ($5.5 \times10^{-3}$) & (0.010) \\[3pt]
Enet & 0.246 & 5.46 & 0.93 & 0.69 \\
& ($29.9 \times10^{-4}$) & (0.081) & (0.013) & (0.032) \\[3pt]
S.enet & \textbf{0.232} & 5.30 & \textbf{0.98} & 0.59 \\
& ($7.6 \times10^{-4}$) & (0.078) & ($7.8 \times10^{-3}$) & (0.029) \\[3pt]
Ada.s.enet & \textbf{0.232} & \textbf{5.25} & 0.91 & \textbf{0.82} \\
& ($15.0 \times10^{-4}$) & (0.075) & ($21.0 \times10^{-3}$) & (0.020)
\\
\hline
\end{tabular*}
\tabnotetext[]{}{For annotation, see Table \protect\ref{tab:blocksetting}.}
\end{table}

\begin{table}
\caption{Results for the block setting, averaged over 50 simulations.
We make use of the following abbreviations: ``PE'' for ``mean squared
prediction error,'' ``g.ridge'' for ``generalized ridge,'' ``enet''
for ``elastic net,'' ``s.enet'' for ``structured elastic net,'' and
``ada.s.enet'' for ``adaptive structured elastic net.'' Standard errors
are given in parentheses. For each column, the best performance is
emphasized in boldface}\label{tab:blocksetting}
\begin{tabular*}{\textwidth}{@{\extracolsep{\fill}}lcccc@{}}
\hline
\textbf{Method} & $\bolds{L^1(\wh{\bolds{\beta}}, \bolds{\beta}^{\ast})}$ & \textbf{PE} &
\textbf{Sensitivity} & \textbf{Specificity} \\
\hline
Ridge & 0.082 & 5.41 & & \\
& ($3.4 \times10^{-3}$) & (0.080) & & \\[3pt]
G.ridge & 0.064 & 5.35 & & \\
& ($1.9 \times10^{-3}$) & (0.078)& & \\[3pt]
P-spline & 0.065 & 5.34 & & \\
& ($1.9 \times10^{-3}$) & (0.077)& & \\[3pt]
Lasso & 0.207 & 6.12 & 0.73 & 0.62 \\
& ($3.6 \times10^{-3}$) & (0.089) & ($7.5 \times10^{-3}$)& (0.014) \\[3pt]
Fused lasso & \textbf{0.058} & 5.34 & \textbf{0.99} & 0.51 \\
& ($1.9 \times10^{-3}$) & (0.076) & ($7 \times10^{-4}$) &
($0.009$) \\[3pt]
Enet & 0.094 & 5.47 & 0.95 & 0.73 \\
& ($5.0 \times10^{-3}$) & (0.072) & ($6.4 \times10^{-3}$) & (0.083) \\[3pt]
S.enet & 0.070 & 5.38 & \textbf{0.99} & 0.60 \\
& ($5.0 \times10^{-3}$) & (0.080) & ($3.3 \times10^{-3}$) & (0.027) \\[3pt]
Ada.s.enet & 0.061 & \textbf{5.32} & 0.97 & \textbf{0.83} \\
& ($3.2 \times10^{-3}$) & (0.69) & ($8.0 \times10^{-3}$) & (0.018) \\
\hline
\end{tabular*}\vspace*{-2pt}
\end{table}

\begin{figure}

\includegraphics{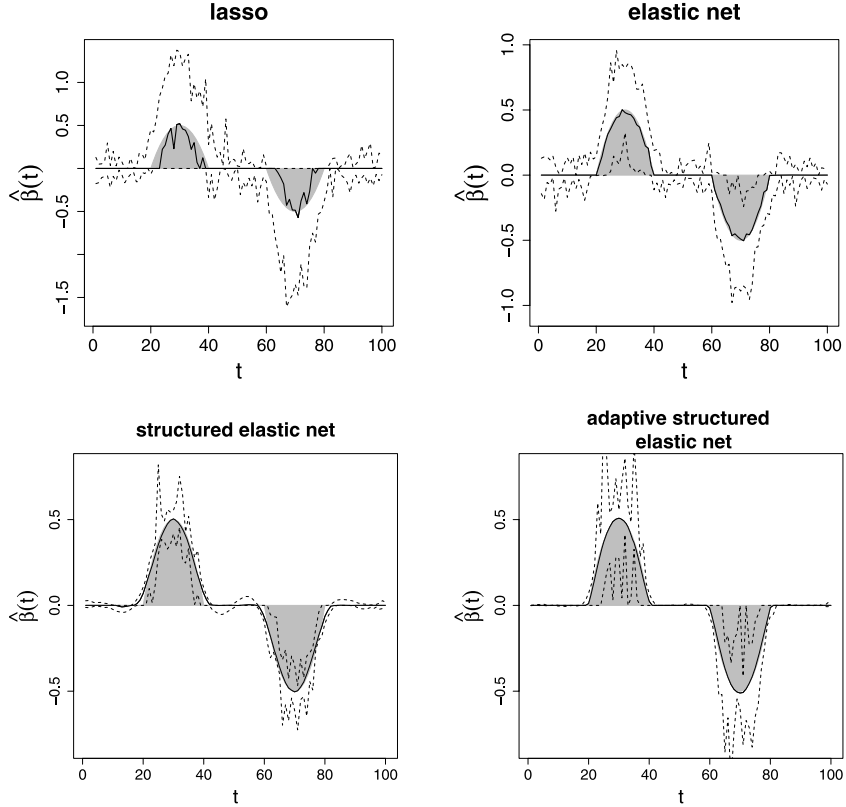}

\caption{Estimated coefficient functions for the bump setting. The pointwise
median curve over 50 iterations is represented by a solid line,
pointwise 0.05- and 0.95-quantiles are drawn in dashed lines.}\label{fig:estcurvesbumps}
\end{figure}

\begin{figure}

\includegraphics{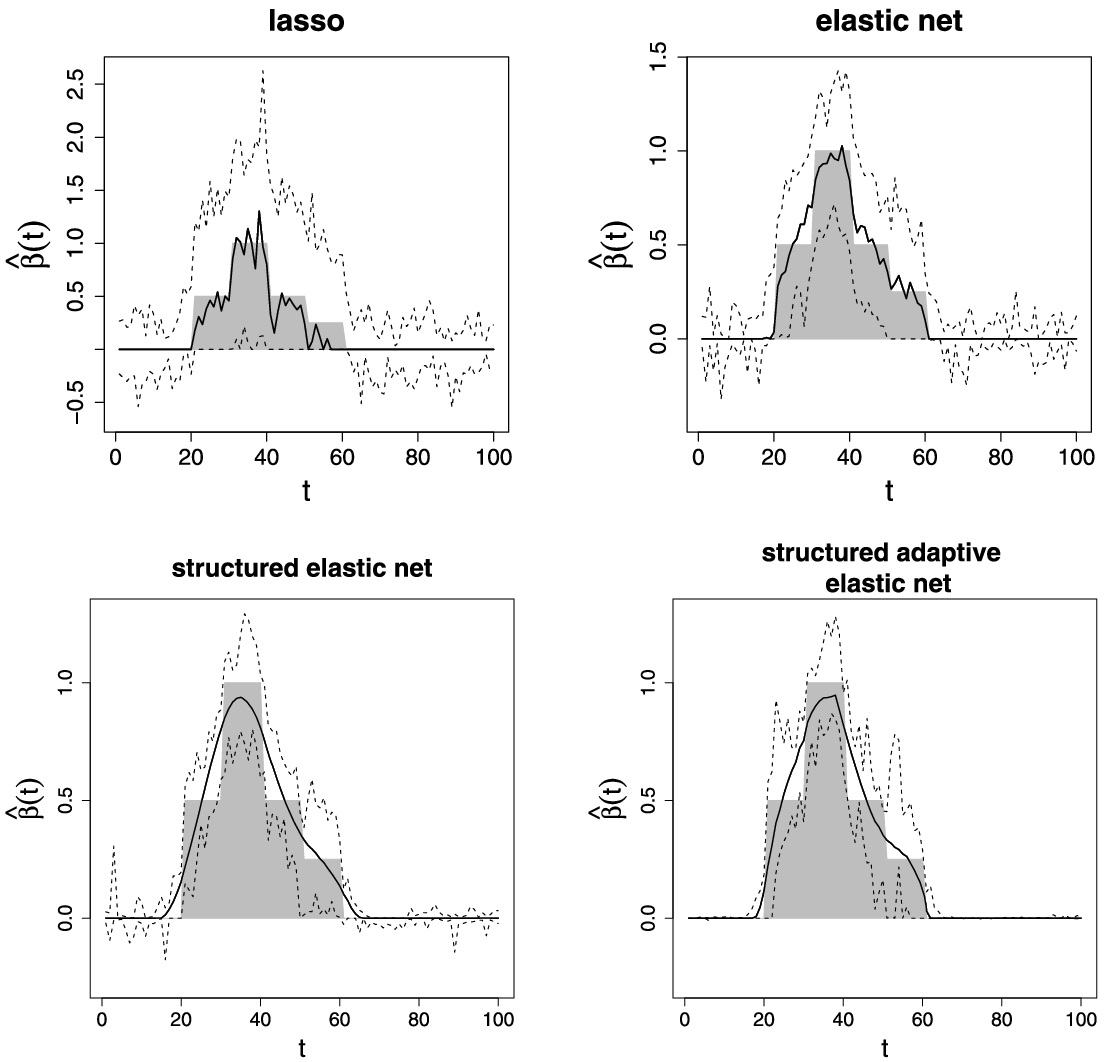}

\caption{Estimated coefficient functions for the block setting.}\label{fig8}
\end{figure}

For the bump setting, Figure \ref{fig:estcurvesbumps} shows that the
double-regularized procedures employing decorrelation clearly
outperform a
visibly unstable lasso. Due to a favorable signal-to-noise ratio, even
simplistic approaches such as ridge- or generalized ridge regression show
competitive performance with respect to prediction of future
observations. In
pure numbers, the estimation of $\beta^{\ast}(t)$ is satisfactory as
well. However, the lack of sparsity results into ``noise fitting'' for those
parts where $\beta^{\ast}(t)$ is zero. For the two settings examined here,
the P-spline approach does not improve over generalized ridge regression,
because the two coefficient functions are not overly smooth. The
elastic net
considerably improves over the lasso, 
but it lacks smoothness. Its numerical inferiority to ridge regression results
from double shrinkage as discussed in \citet{Zhou2005}. The
performance of the structured elastic net is not fully satisfactory. In
particular, at the changepoints from zero- to nonzero parts, there is a
tendency to widen unnecessarily the support of the nonzero sections. This
shortcoming is removed by the adaptive structured elastic net, thereby
confirming the theoretical result concerning selection consistency. This
quality seems to be supported by the eminent performance with respect to
sensitivity and specificity. The success of the adaptive strategy is
also founded
on the good performance of the ridge estimator providing the component-specific
weights $\omega(t)$. The block setting is actually tailored to
the fused lasso, whose output are piecewise constant coefficient
functions. Nevertheless, it is not optimal, as the shrinkage of the
$\ell^1$-penalty acts on all coefficients, including those different from
zero. As a result, the fused lasso is outperformed by the adaptive structured
elastic net with respect to prediction, though the structure part is
seen to be
not fully appropriate in the block setting (cf. Figure \ref{fig8}). As opposed to the bump
setting, fitting the
block function seems to be much more difficult to accomplish in general.

\subsubsection{Accelerometer data}\label{sec6.1.2}

\begin{figure}

\includegraphics{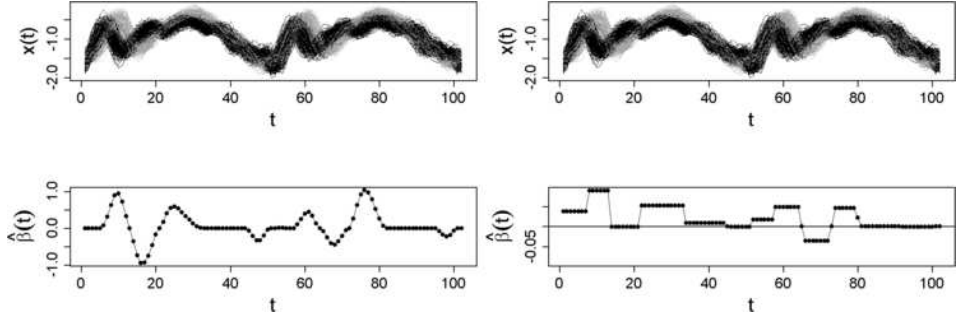}

\caption{Coefficient functions for structured elastic net-regularized
logistic regression (left panel) and the fused lasso support vector machine
(right panel). Within each panel, the upper panel displays the overlayed
double step patterns of the complete sample (406 double steps). The
colors of
the curves refer to the two persons.}\label{fig:accelerometer1} 
\end{figure}

The ``Sylvia Lawry Centre for Multiple Sclerosis Research e.V.,'' Munich, kindly
provided us with two accelerometer records of two healthy female
persons, aged
between 20 and 30. They were equipped with a belt containing an accelerometer
integrated into the belt buckle before walking several minutes on a
flat surface at
a moderate speed. The output are triaxial (vertical, horizontal, lateral)
acceleration measurements at roughly 25,000 sampling points per
person. Following \citet{Daumer2007}, human gait, if defined as the
temporal evolution of
three-dimensional accelerations of the center of mass of the body, is supposed
to be a quasi-periodic process. Every period defines one gait cycle/double
step, which starts with the heel strike and ends with the heel strike
of the
same foot. A single step ends with the heel strike of the other
foot. Therefore, a double step can be seen as a natural unit. As a
consequence, decomposition of the raw signal into pieces, each representing
one double step, is an integral part of data preprocessing, not
described in further detail here. Overall, we extract $i=1,\ldots
,n=406$ double steps, $242$ from person
B ($y = 0$) and $164$ from person A ($y=1$), ending up with a sample
$\{(\mathbf{x}_i, y_i)\}_{i=1}^n$, where each $\mathbf{x}_i = (x_i(t)),
t=1,\ldots,T=102,$ stores the observed vertical acceleration
within double step $i,     i=1,\ldots,n$. For simplicity, we neglect
the dependence of
consecutive double steps within the same person and treat them as independent
realizations. Horizontal- and lateral acceleration are not considered, since
they do not carry information relevant to our prediction problem. We
aim at
the prediction of the person (A or B) given a double step pattern, and
additionally at the detection of parts of the signal apt for discriminating
between the two persons. We randomly divide the complete sample into a
learning set of size 300 and a test set of size 106, and subsequently
carry out logistic
regression on the training set, using the structured elastic net with a squared
first difference penalty. Hyperparameters are determined by ten-fold
cross-validation, and the resulting logistic regression model is used to
obtain predictions for the test set. The fused lasso with the hinge
loss of support
vector machines is used as competitor. A collection of results is
assembled in
Figure \ref{fig:accelerometer1} and Table \ref{tab:accelerometer},
from which one concludes that
classification is an easy task, since (nearly) perfect
misclassification rates
on the test set are achieved. Concerning feature selection, the results
of the
structured elastic net are comparable to those of the fused lasso.

\begin{table}[b]
\caption{Results of step classification for the fused lasso support vector
machine and structured elastic net-regularized logistic regression. The
bound imposed on the 1-norm of
$\bolds{\beta}$ corresponding to $\lambda_1$ is denoted by $t_1$, while $t_2$
corresponding to $\lambda_2$ denotes the bound imposed on
the absolute differences $\sum_{t=2}^T |\beta(t) - \beta(t-1)|$ for the
fused lasso and the squared differences $\sum_{t=2}^T (\beta(t) -
\beta(t-1))^2$ for the structured elastic net, respectively.
Concerning the degrees of
freedom of the two procedures, we take the number of nonzero blocks for the
fused lasso. For the structured elastic net, we make use~of~a heuristic due
to Tibshirani (\protect\citeyear{Tib1996}) that rewrites
the lasso fit as the weighted ridge
fit; see~Slawski, zu Castell and Tutz (\protect\citeyear{Sla2009}) for details}\label{tab:accelerometer}
\begin{tabular*}{\textwidth}{@{\extracolsep{\fill}}lcccc@{}}
\hline
$\bolds{t_1}$ & $\bolds{t_2} $ & \textbf{Test error} & \textbf{Degrees of freedom} & \textbf{\# nonzero coefficients}\\
\hline
\multicolumn{5}{@{}c@{}}{Fused lasso}\\
\hphantom{0}2.5 & 0.5 & 0 & 9\hphantom{00.} & 46 \\[3pt]
\multicolumn{5}{@{}c@{}}{Structured elastic net} \\
23 & 2\hphantom{0.} & 1 & 7.85 & 61 \\
\hline
\end{tabular*}
\end{table}
%

\subsection{Surface fitting}\label{sec6.2}

\begin{figure}

\includegraphics{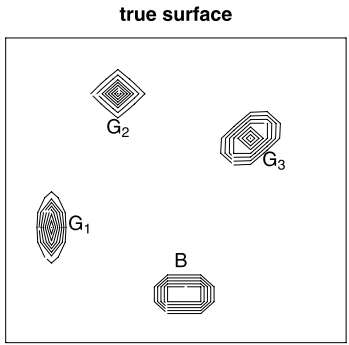}

\caption{Contours of the surface according to equation (\protect\ref{eq:betasurface}).}\label{fig:truesurface}
\end{figure}

Figure \ref{fig:truesurface} depicts the surface to be fitted on a $20
\times
20$ grid. The surface can be represented by a discrete function
$\beta^{\ast}(t,u),     t,u=1,\ldots,20$. It consists of three
nonoverlapping truncated Gaussians of different shape and one plateau
function. We have
\begin{eqnarray}\label{eq:betasurface}
\qquad \beta^{\ast}(t,u) &=& B(t,u) + G_1(t,u) + G_2(t,u) + G_3(t,u), \nonumber\\
B(t,u) &=& \frac{1}{2} I(t \in\{10,11,12 \}, u \in\{3,4 \}), \\
G_1(t,u) &=& \max \left\{0, \exp \left(-(t-3     u-8)  \pmatrix{
3 & 0 \cr
0 & 0.25
}  \pmatrix{
t - 3 \cr
u - 8 }
 \right) - 0.2  \right\},\nonumber\\
G_2(t,u) &=& \max \left\{0, \exp \left(-(t-7     u-17)\vphantom{\pmatrix{
0.5 & -0.25 \cr
-0.25 & 0.5
}}\right.\right.\nonumber\\
&&\left.\left.\hphantom{ \max \left\{0, \exp \left(\vphantom{\pmatrix{
0.5 & -0.25 \cr
-0.25 & 0.5
}}\right.\right.}
{}\times   \pmatrix{
0.75 & 0 \cr
0 & 0.75
}\pmatrix{
t - 7 \cr
u - 17 }\right) - 0.2  \right\},\nonumber \\
G_3(t,u) &=& \max \left\{0, \exp \left(-(t-15     u-14)\vphantom{\pmatrix{
0.5 & -0.25 \cr
-0.25 & 0.5
}}\right.\right.\nonumber\\
&&\left.\left.\hphantom{ \max \left\{0, \exp \left(\vphantom{\pmatrix{
0.5 & -0.25 \cr
-0.25 & 0.5
}}\right.\right.}
{}\times  \pmatrix{
0.5 & -0.25 \cr
-0.25 & 0.5
}  \pmatrix{
t - 15 \cr
u - 14 }
\right) - 0.2  \right\}.\nonumber
\end{eqnarray}

\begin{figure}

\includegraphics{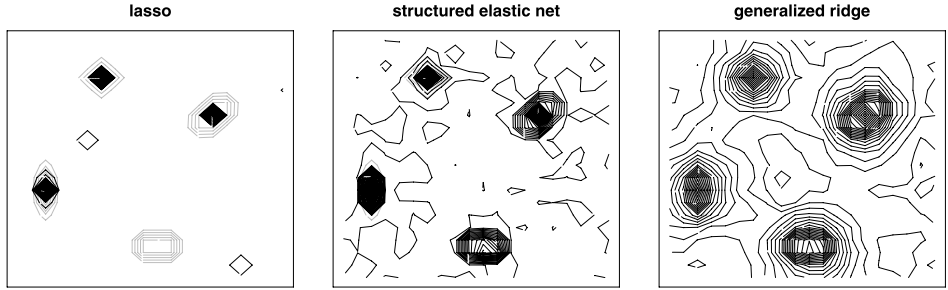}

\caption{Contours of the estimated surfaces for three selected methods,
averaged pointwise over 50 runs.}\label{fig:surfaces}
\end{figure}

Similarly to the simulation study in Section \ref{sec6.1.1}, we simulate a noisy
version of the surface according to
\[
y(t,u) = \beta^{\ast}(t,u) + \epsilon(t,u), \qquad \{ \epsilon(t,u)\}
\stackrel{\mathrm{i.i.d.}}{\sim} N(0, 0.25^2), \  t,u=1,\ldots,20.
\]
For each of the 50 runs, we simulate two instances of $y(t,u)$. The first
one is used for training and the second one for hyperparameter tuning.
The mean squared error
for estimating $\bolds{\beta}^{\ast}$ is computed and averaged over 50
runs. Results are summarized in Figure \ref{fig:surfaces} and Table
\ref{tab:surfaces}. We compare ridge, generalized ridge with a difference
penalty according to the grid structure, lasso, fused lasso with a total
variation penalty along the grid, structured- and
adaptive structured elastic net with the same difference penalty as for
generalized ridge. The elastic net coincides---up to a constant scaling
factor---with the lasso/soft thresholding in the orthogonal design case and
is hence not considered.

\begin{table}[b]
\tablewidth=310pt
\caption{Results of the simulation, averaged over 50 iterations (standard
errors in parentheses). The columns labeled $B$, $G_1$, $G_2$, $G_3$, and
``zero'' contain the mean prediction error for the corresponding region
of the surface. The
abbreviations equal~those in Table \protect\ref{tab:blocksetting}. The
prediction error
has been rescaled by 100}\label{tab:surfaces}
\begin{tabular*}{310pt}{@{\extracolsep{\fill}}lcccccc@{}}
\hline
\textbf{Method} & \textbf{PE} & $\bolds{B}$ & $\bolds{G_1}$ & $\bolds{G_2}$ & $\bolds{G_3}$ & \textbf{Zero} \\
\hline
Ridge & 1.20 & 0.31 & 0.28 & 0.20 & 0.34 & 0.07 \\
& (0.01) & & & & & \\[3pt]
G.ridge & 1.17 & 0.18 & 0.20 & 0.14 & 0.21 & 0.44 \\
& (0.04) & & & & & \\[3pt]
Lasso & 1.31 & 0.37 & 0.32 & 0.22 & 0.39 & \textbf{0.01} \\
& (0.01) & & & & & \\[3pt]
Fused lasso & 0.67 & \textbf{0.14} & 0.12 & \textbf{0.08} & \textbf{0.15} &
0.18 \\
& (0.02) & & & & & \\[3pt]
S.enet & 0.88 & 0.22 & 0.16 & 0.12 & 0.23 & 0.18 \\
& (0.02) & & & & & \\[3pt]
Ada.s.enet & \textbf{0.56} & 0.15 & \textbf{0.09} & \textbf{0.08} & 0.18
& 0.06 \\
& (0.02 )& & & & & \\
\hline
\end{tabular*}
\end{table}

\section{Discussion}\label{sec7}

The structured elastic net is proposed as a procedure for coefficient
selection and smoothing. We have
established a general notion of structured features, for which the structured
elastic net is able to take advantage of prior knowledge as opposed to
the lasso and
the elastic net, which are both purely data-driven. The structured elastic
net may also be regarded as a computationally more convenient
alternative to
the fused lasso. Conceptually, generalizing the fused lasso by
computing the
total variation of the coefficients along a graph is straightforward. However,
due to the nondifferentiability of the structure part of the fused lasso,
computation may be intractable even for moderately sized graphs.

Turning to the drawbacks of the structured elastic net, it is obvious
that model selection and computation of standard
errors and, in turn, the quantification of uncertainty, are notoriously
difficult.
A Bayesian approach promises to be superior in this regard. The
lasso can be treated within a Bayesian inference framework [\citet
{Casella2008}],
while the quadratic part of the structured elastic net regularizer is already
motivated from a Bayesian perspective in this paper.

With regard to possible directions of future research, we will consider
studying the
structured elastic net in combination with other loss functions, for
example, the hinge loss of support vector machines
or the check loss for quantile regression. 
The asymptotic analysis in this paper is basic in the sense that it is
bound to strong assumptions, and the role
of the structure part of the regularizer and its interplay with the true
coefficient vector is not well understood yet, leaving some room for more
profound investigations.

\section*{Acknowledgments}

We thank the Sylvia Lawry Centre Munich e.V.
for its support with the accelerometer data example, in particular,
Martin Daumer for numerous discussions, and Christine Gerges and
Kathrin Thaler
for producing the data. We thank Jelle Goeman for one helpful
discussion about
his algorithm and making his code publicly available as \texttt{R} package.
We are grateful to Angelika van der Linde and Daniel Saban\'es-Bov\'e for
pointing us to several errors and typos in earlier drafts.

We thank two reviewers, an associate editor, and an area
editor for their constructive comments and suggestions, which helped us to
improve on an earlier draft.

\begin{supplement}
\stitle{Supplement to ``Feature Selection guided by Structural Information''}
\slink[doi]{10.1214/09-AOAS302SUPP}
\slink[url]{http://lib.stat.cmu.edu/aoas/302/supplement.pdf}
\sdatatype{.pdf}
\sdescription{The supplement contains proof of all statements of the
main article.}
\end{supplement}

\printaddresses

\end{document}